\documentclass[article]{IEEEtran}
\usepackage[utf8]{inputenc}
\usepackage{xcolor}
\usepackage{blindtext, graphicx}
\usepackage{amsmath}
\usepackage{textcomp}
\usepackage{graphicx} 
\usepackage{amsmath,amssymb}  
\usepackage{paralist}
\usepackage[shortlabels]{enumitem}
\usepackage{bm}  
\usepackage{multirow}
\usepackage[caption=false,font=footnotesize]{subfig}
\usepackage[pdftex,bookmarks,colorlinks=false,breaklinks,hidelinks]{hyperref}  
\usepackage{memhfixc} 
\usepackage[mathscr]{euscript}
\DeclareSymbolFont{rsfs}{U}{rsfs}{m}{n}
\DeclareSymbolFontAlphabet{\mathscrsfs}{rsfs}
\title{Optimizing Grid Resilience: A Capacity Reserve Market for High Impact Low Probability Events
\thanks{U.~T.~Salman and Z.~Wang are with the Department of Electrical and Computer Engineering at the University of Connecticut, Storrs, Connecticut, USA 06268.  ($^{\dagger}$corresponding author: Zongjie Wang, zongjie.wang@uconn.edu). }%
\thanks{T. M. Hansen is with Electrical Engineering and Computer Science Department, South Dakota State University, Brookings, South Dakota, USA 57007.}%
\thanks{This work was funded in part by the Eversource project entitled ``A Pathway to Enhance Grid Resilience: Zero-Carbon Energy Communities with DER-based ELCC Quantification.''}%
}

\author{Umar Taiwo Salman, Zongjie Wang$^\dagger$, Timothy M. Hansen\vspace{-24pt}}
\usepackage{cite}
\usepackage{graphicx}
\usepackage{nomencl}
\makenomenclature

\providetoggle{nomsort}
\settoggle{nomsort}{true} 

\makeatletter
\iftoggle{nomsort}{%
    \let\old@@@nomenclature=\@@@nomenclature        
        \newcounter{@nomcount} \setcounter{@nomcount}{0}%
        \renewcommand\the@nomcount{\two@digits{\value{@nomcount}}}
        \def\@@@nomenclature[#1]#2#3{
          \addtocounter{@nomcount}{1}%
        \def\@tempa{#2}\def\@tempb{#3}%
          \protected@write\@nomenclaturefile{}%
          {\string\nomenclatureentry{\the@nomcount\nom@verb\@tempa @[{\nom@verb\@tempa}]%
          \begingroup\nom@verb\@tempb\protect\nomeqref{\theequation}%
          |nompageref}{\thepage}}%
          \endgroup
          \@esphack}%
      }{}
\makeatother

\begin{document}
\maketitle
\begin{abstract}
     This paper addresses the challenges of high-impact low-probability (HILP) events by proposing a novel capacity reserve event market for mobile generation assets, aimed at supporting the transmission network during such incidents. Despite the usefulness of portable generators and mobile energy units in restoring power, there are drawbacks such as environmental impact, finite operation, and complex cost recovery. The proposed market integrates these resources into a dispatch framework based on pre-established contracts, ensuring fair compensation and considering factors like capacity, pricing, and travel distance. Resource owners receive advanced notifications for potential events, allowing them to adjust their bids for cost recovery. Simulations on an IEEE 30-bus case have been conducted to demonstrate the model effectiveness in increasing grid resiliency.
\end{abstract}

\vspace{-15 pt}
\begin{IEEEkeywords}

Event market, extreme weather, HILP events, resilience, mobile generating resources.
\vspace{-15pt}
\end{IEEEkeywords}

\nomenclature{$d$ \in \(\mathscr{D}\)}{Index of the duration of event}
\nomenclature{\(\rho\) \in \(\Omega\)}{Index of scenario generation}
\nomenclature{\(r\)}{System reserve percent}
\nomenclature{\(P^D\)(\(\pi^D\))}{Quantity (price) in \$/MWh}
\nomenclature{\(t\) \in \(T\)}{Index of time duration for the period T}
\nomenclature{\(j\) \in \(\mathscr{J}\)}{Index of network area}
\nomenclature{\(q_1, q_2, ...\) \in \({N_Q}\)}{Index of buses}
\nomenclature{\(b\) \in \(N_B\)}{Block of storage participating in capacity offers}
\nomenclature{\(v\) \in \(N_V\)}{Block of EVs participating in capacity offers}
\nomenclature{\(g\) \in \(N_G\)}{Block of generators participating in capacity offers}
\nomenclature{\(\pi^{ESR}\) (\(E^{C,ESR}\))}{ESR price (quantity) capacity offers}
\nomenclature{\(\pi^{EV}\) (\(E^{C,EV}\))}{EVs price (quantity) capacity offers}
\nomenclature{\(\pi^{G}\) (\(\overline{P}^{G}\))}{Generator price (quantity) offered}
\nomenclature{\(\pi^{G,RS}\) (\(\overline{P}^{G,RS}\))}{Generator price (quantity) spinning reserve offers}
\nomenclature{\(\psi\)}{Constant degradation capacity-dependent cost}
\nomenclature{\(P^{Loss}\)}{Transmission loss}
\nomenclature{\(J\)}{Objective function}
\nomenclature{\(\Gamma\)}{Fraction of the quantity share of ESR (EV) energy capacity available for discharge}
\nomenclature{\(\Gamma^{RS}\)}{Fraction of the quantity share of ESR (EV) energy capacity available for capacity reserve}
\nomenclature{\(Y^G\) (\(Y^{G,RS}\))}{Binary variable indicating whether a generator (reserve) offer is cleared or not \in [1,0]}
\nomenclature{\(x^{ESR}\) (\(x^{EV}\))}{Binary variable indicating whether a ESR (EV) offer is discharging or not \in [1,0]}
\nomenclature{\(X^{ESR}\) (\(X^{EV}\))}{Binary variable indicating whether a ESR (EV) capacity offer is cleared or not \in [1,0]}
\nomenclature{\(Z^{D}\)}{Binary variable indicating whether an outage demand exists or not \in [1,0]}
\nomenclature{\(\overline{SoC}\)}{Maximum offer quantity of the ESR (EV) in p.u.}
\nomenclature{\(\underline{SoC}\)}{Minimum offer quantity of the ESR (EV) in p.u.}
\nomenclature{\(SoC^{C,RS}\)}{Cleared capacity of ESR (EV) in MWh}
\nomenclature{\(SoC^{RS}\)}{Cleared capacity of ESR (EV) in p.u.}
\nomenclature{\(\overline{P}\) (\(\underline{P}\))}{Maximum (Minimum) limit on generator generator power output in MW}
\nomenclature{\(\chi\)}{Distance traveled by generating resources in km}
\nomenclature{\(\pi^\chi\)}{Cost of shipping a resource from its initial location to the deployed location \$/km MWh}

\printnomenclature
\vspace{-10pt}
\section{Introduction}
\label{intro}

This paper proposes a novel capacity reserve event market tailored for mobile generating systems, e.g., electric vehicles (EVs), addressing high impact low probability (HILP) events to facilitate the deployment of resources upon request by grid operators, according to pre-agreed terms. The central strategy involves a grid-side model (GSM) that factors in each resource's unique attributes are compensated judiciously for enhanced grid resilience during HILP scenarios. 
Resource owners in this model are proactively notified about potential events and their associated probabilities, enabling each resource to submit offers based on specific characteristics like degradation, capacity, duration, and travel distance, along with other factors essential for cost recovery. The establishment of this event market significantly reduces outage times by economically sourcing power from alternative resources to replenish lost energy. Another key advantage of this market design is its ability to prevent system fragmentation in the downstream network, which avoids unnecessary microgrid formations in areas not directly affected by the disaster, but still dependent on the impacted generators. This approach ensures a more resilient grid and facilitates a more efficient and cost-effective energy distribution in times of crisis.

The concept of resilience, viewed from multiple perspectives, is fundamentally about a system's capacity to foresee, adjust to changing conditions, and quickly rebound after disturbances \cite{resiliencepesgm}. This is particularly relevant in the context of physical electric delivery networks, which are often susceptible to severe disruptions caused by natural events like hurricanes, typhoons, or earthquakes. For instance, the extensive damage inflicted by such events can significantly impact communities, often requiring weeks or even months for complete repairs – a situation exemplified by Hurricane Maria in Puerto Rico, where restoring power to all customers took nearly a year \cite{kwasinski2019hurricane}. The proposed market model addresses this need for resilience by enabling a more agile and responsive power grid, capable of minimizing the impact of these events and facilitating a faster return to normal operations. Through this lens, the study's focus on enhancing grid resilience via a novel market strategy becomes a vital component of the broader efforts to strengthen the power system's reliability and robustness against unforeseen challenges. 

Resilience research focusing on optimizing resources for system disruptions in distribution networks and microgrids is increasing. Techniques such as stochastic scheduling are used to lessen uncertainty impacts on grid operations, yet achieving resilience in power networks is challenging \cite{ZW,chen2017modernizing}. Integrating distributed energy resources (DERs) into microgrids is identified as a key solution for grid resilience against extreme events \cite{YOUNESI2022112397,li2022coordinating,gargari2023preventive}, although this can increase scheduling costs, necessitating efficient resource management to control expenses. Recent research has delved into innovative strategies to overcome resilience complexities, focusing on advanced planning and operational mechanisms to overcome these challenges. In \cite{yao2020integrated}, a strategy for improving grid resilience through transmission expansion and black start allocation using battery energy storage systems (BESS) is proposed, aiming to reduce power generation costs and load shedding in disasters. A BESS-based framework is introduced in \cite{nazemi2019energy} for energy storage planning and enhancing distribution grid resilience and emergency response for HILP events.  The BESS role in resilience during hurricanes, addressing challenges like resource scheduling, cost recovery, and strategic network placement, is examined in \cite{nguyen2019assessing}.

Despite advancements in grid resilience via strategies like stochastic scheduling and planning, unresolved challenges persist in resource scheduling, cost recovery modeling, and strategic resource placement. This underscores the need for innovative approaches in managing HILP events. 
This paper addresses timely and efficient resource scheduling challenges via a novel HILP event market  that incorporates resource-specific characteristics and travel distances in the dispatch process, ensuring improved benefits for both participating resources and the grid operator. The main contributions of this work are: (i) designing a market model for efficient use of mobile generators in emergencies; and (ii) developing a resiliency compensation system accounting for capacity, travel, and resource specifics. The method is showcased on the the IEEE 30-bus test case, highlighting the market's ability to enhance grid resilience and reliability with strategic resource scheduling and dispatch in crisis situations, and reducing outages through optimal resource management.

\section{Problem Formulation}
\label{formulation}
\subsection{The Event Market}

\begin{figure}[!htb]
    \centering
    \includegraphics[width=1\linewidth]{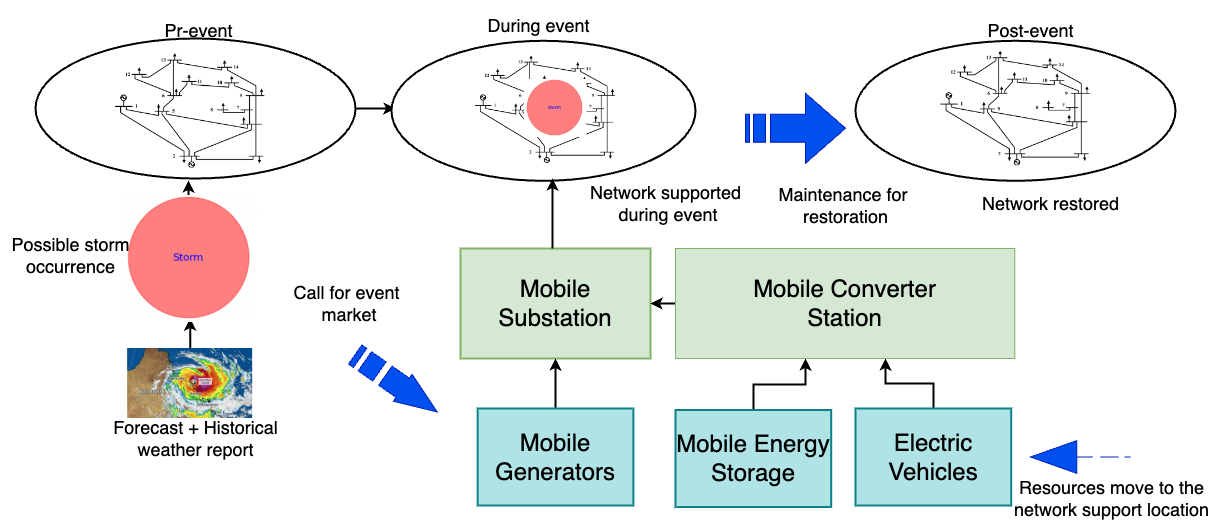}
    \caption{Mobile resources are activated in response to a HILP event. The operator activates the event market, alerting participants and coordinating the dispatch of resources to designated locations.}
    \label{fig:schematics}
    \vspace{-12pt}
\end{figure}

This paper proposed an advanced GSM to improve grid resilience against HILP events. Utilizing historical and forecast data, system operators assess storm probabilities and activates an event market (see Fig. \ref{fig:schematics}) as needed. Resource providers submit offers with capacities and prices to enable proactive and strategic response to events.

The GSM is designed to strategically reserve sufficient capacity from diverse sources prior to the forecast events using the submitted offers to clear the market efficiently. This approach ensures not only the rapid mobilization of resources, but also their optimal allocation based on the network's anticipated outage profile. Unique to this model is the integration of  physical resource characteristics, ensuring that all associated costs, including physical deployment and operational costs, are factored into the market clearing process. This dynamic strategy facilitates a more responsive and efficient deployment of resources, substantially minimizing network damage and curtailing system downtime during such events. Therefore, the GSM contributes to a significant increase in the system's overall resilience. The model's innovative approach to market clearing and resource allocation, with a keen focus on cost-effectiveness and operational efficiency, marks a substantial advancement in the field, showcasing a novel way to enhance grid resilience in the face of unpredictable and destructive HILP events.
\vspace{-12pt}
\subsection{Data \& Market Terms Description}
 
Following the overview of the event market, this subsection details the data and market terms pertinent to mobile power-generating resources such as energy storage resources (ESRs), EVs, and distributed generators (DGs). Critical terms and parameters for the event market operation include:
\begin{itemize}
    \item \textbf{Outages}: The grid operator estimates the total capacity outage, denoted as $\overline{P}_{D}$, at an estimated price $\pi^D$, typically resulting from windstorm impacts. This represents the total energy capacity available for bidding in the event market.
    \item \textbf{Generator offer:} Generators propose a supply quantity $\overline{P}^G$ at a specified price $\pi^G$.  
    \item \textbf{ESR Offer:} ESRs submit a fractional offer of their capacity, $\Gamma^{ESR}$, based on $E^{C,ESR}$, at a price $\pi^{ESR}$.
    \item \textbf{EV Offer: } EVs make a similar fractional offer, $\Gamma^{EV}$, of their capacity $E^{C,EV}$, priced at $\pi^{ESR}$.

        \item \textbf{Spinning reserve offer:} DGs provide a spinning reserve quantity $\overline{P}^{G,RS}$ at price $\pi^{G,RS}$. ESRs offer a spinning reserve fraction $\Gamma^{RS}$, indicating the ESR capacity portion available for spinning reserve at price $\pi^{EC,RS}$. Notably, EVs do not participate in the spinning reserve market. 
        \end{itemize}
        
\hspace{0.1em}        
The decision variables include the cleared capacities for DGs, ESRs, and EVs, as well as the cleared spinning reserve capacities for DGs and ESRs, in addition to the associated binary variables as defined in the nomenclature.

\vspace{-15pt}
\subsection{Grid-Side Modeling}
\label{modeling1}
The objective $J$ of the event market, defined in \eqref{eq:jrss},  is to maximize the adequacy of the scheduling resources participating in the event market leading to high benefit for all participants (i.e., improved social welfare):
\begin{equation}
\small
\label{eq:jrss}
    J = C_{OT} - C_{SRC} - C_{Sch} - C_{TRN}
    \vspace{-16pt}
\end{equation}

\begin{equation}
\small
\label{eq:outagecost}
    C_{OT=}\sum_{\rho}^{\Omega} \sum_d^{\mathscr{D}} \sum_j^\mathscr{J} \sum_h^{N_D} \pi^D_{h,j,d,\rho} P^D_{h,j,d,\rho} 
     \vspace{-15pt}
\end{equation}
\begin{gather}
\small
\label{eq:spinningcost}
    C_{SRC} = \sum_{\rho}^{\Omega} \sum_d^{\mathscr{D}} \sum_j^\mathscr{J} \Big(\sum_b^{N_\mathscr{B}} \pi^{ESR}_{b,j,d,\rho}  SoC^{C,RS,ESR}_{b,j,d,\rho} \\\nonumber 
   +\sum_g^{N_G}\pi^{G,RS}_{g,j,d,\rho} P^{G,RS}_{g,j,d,\rho}\Big)
\end{gather}
\vspace{-2.5pt}
\begin{gather}
\small
   C_{Sch}=\sum_{\rho}^{\Omega} \sum_t^{\mathscr{T}} \sum_j^\mathscr{J} 
   \Big( \sum_{b/v}^{N_\mathscr{B/V}} \Big ( \alpha_{b/v} SoC^{dch}_{b/v,j,t,\rho} \nonumber \\+ \beta_{b/v} SoC^{dch}_{b/v,j,t-1,\rho} + \gamma_{b/v} P^{ESR}_{b/v,j,t,\rho} + \psi_{b/v} \Big) \nonumber \\
+  U_{j,t,\rho} \pi^{su}_{j,t,\rho}+ V_{j,t,\rho} \pi^{sd}_{j,t,\rho} + \sum_g^{N_G} \pi^G_{g,j,t,\rho} P^G_{g,j,t,\rho} \Big)\Big) 
\label{eq:schedulecosts}
  \end{gather}

\vspace{-10pt}
  \begin{gather}
  \small
 C_{TRN} =  \sum_{\rho}^{\Omega} \sum_t^{\mathscr{T}} \sum_j^\mathscr{J} \Big( \sum_b^{N_\mathscr{B}} \pi^\mathscr{\chi}_{b,j,d,\rho} \chi^{ERS}_{b,j,d,\rho} EC^{C,ESR}_{b,j,d,\rho}  \nonumber \\ +\sum_{v}^{N_\mathscr{V}} \pi^\mathscr{\chi}_{v,j,d,\rho} \chi^{EV}_{v,j,d,\rho} EC^{C,EV}_{v,j,d,\rho}+\sum_g^{N_G} \pi^\mathscr{\chi}_{g,j,d,\rho} \chi^{G}_{g,j,d,\rho}P^{G}_{g,j,d,\rho}\Big) \Big)\label{eq:transportcost}
 \vspace{-5pt}
\end{gather}
where $C_{OT}$, defined in \eqref{eq:outagecost}, is the estimated outage cost; $C_{SRC}$, defined in \eqref{eq:spinningcost}, is the spinning reserve capacity cost for DGs and ESR; $C_{Sch}$, defined in \eqref{eq:schedulecosts}, is the cost associated with ESR/EV for energy supply during discharging considering
degradation influenced by depth of discharge (DOD), discharge rate, and the
overall cost of generators (i.e., expenses related to generator start-up, shutdown, and energy consumption); and $C_{TRN}$, defined in \eqref{eq:transportcost}, is the cost of moving the resources from their initial
location to deployed areas.
\vspace{-12pt}
\subsection{Balance Constraint}
\vspace{-5pt}
\label{constraint1}
This constraint ensures the total supply from generating resources matches the estimated outage quantity, as dictated by the DC power flow equations detailed in \eqref{eq:const1}.

\vspace{-15pt}
\begin{gather}
\small
    \sum_{\rho}^{\Omega} \sum_t^{\mathscr{T}} \sum_j^\mathscr{J} \Big(P^D_{j,t,\rho} +  
    P^{Loss}_{j,t,\rho}  \nonumber \Big)\\ 
    = \sum_{\rho}^{\Omega} \sum_t^{\mathscr{T}} \sum_j^\mathscr{J} 
     \Big(\sum_b^{N_\mathscr{B}}   P^{ESR}_{b,j,t,\rho} +  \sum_v^{N_\mathscr{V}} P^{EV}_{v,j,t,\rho} +  \sum_g^{N_{G}} P^{G}_{g,j,t,\rho} \Big)\label{eq:const1} 
\end{gather}

\vspace{-20pt}
\subsection{Network spinning reserve constraint}
\label{constraint2}
Letting the network spinning reserve factor be $r$ percent of the estimated quantity of the network \cite{reddy2017estimation}, then 
\begin{gather} 
\sum_{\rho}^{\Omega} \sum_t^{\mathscr{T}} \sum_j^\mathscr{J} P^D_{j,t,\rho} 
\leq \nonumber\\ (1+r) \Big( \sum_{\rho}^{\Omega} \sum_t^{\mathscr{T}} \sum_j^\mathscr{J} \Big( \sum_b^{N_\mathscr{B}}   P^{ESR}_{b,j,t,\rho} +  \sum_v^{N_\mathscr{V}} P^{EV}_{v,j,t,\rho} +  \sum_g^{N_{G}} P^{G}_{g,j,t,\rho} \Big) \Big) \label{eq:const3} 
\end{gather}
where 
\begin{gather}
\small
    P^{G,RS}_{g,t} \leq \overline{P}_{g,t}-P^G_{g,t,} ~~~~ \forall g \in N_G, t \in \mathscr{T}
\end{gather}
and cleared capacity reserve from an ESR/EV then becomes 
\begin{gather}
    P_{b/v,d} = SoC^{RS}_{b/v,d} E^C_{e/v}, ~~ \forall b \in N_B, \forall v \in N_V, \forall d \in \mathscr{D}.
\end{gather}
\vspace{-25pt}
\subsection{Generator Market Constraints}
\label{constraint3}
The constraints guarantee that the offers from resources in terms of prices and quantities, along with the system's estimations, stay within their maximum limits.

\vspace{-15pt}
\begin{gather}
\small
            P^{G,RS}_{g,j,t,\rho} \leq \Bar{P}^{G,RS}_{g,j,t,\rho} Y^{G,RS}_{g,j,t,\rho}\\
            P^{D,est}_{j,t,\rho} \leq \Bar{P}^{D,est}_{j,t,\rho} Z^{D,est}_{j,t,\rho}\\
            P^{G}_{g,j,t,\rho} \leq \Bar{P}^{ESR}_{g,j,t,\rho} Y^{G}_{g,j,t,\rho}
\end{gather}
\vspace{-28pt}
\subsection{ESRs and EVs Capacity Reserve Constraints}
\label{constraint4}
This constraint ensures that the capacity of the ESR/EV cleared in the event market does not surpass their maximum offer in the resource scheduling stage.

\begin{gather}
\small
    SoC_{b,j,d}^{ESR,RS}  \leq \Gamma^{ESR}_{b,j,d}X^{ESR}_{b,j,d})\overline{SoC}^{ ESR}\\
    SoC_{v,j,d}^{ESR,RS}  \leq \Gamma^{EV}_{v,j,d}X^{EV}_{v,j,d})\overline{SoC}^{EV} 
\end{gather}

Cleared value for ESR and EV in the capacity market is given by,
\begin{gather}
\small
    SoC_{b,d,\rho}^{ESR,RS}E^{C,ESR}_b  - (1-X^{ESR})~\overline{SoC}^{ESR}_{b,d,\rho}~E^{C,ESR}_b \nonumber \\\leq SoC^{C,RS,ESR}_{b,d,\rho}\nonumber \\
   \leq  SoC_{b,d,\rho}^{ESR,RS}~E^{C}_b  - (1-X^{ESR})~\underline{SoC}^{ESR}_{b,d,\rho}~E^{C,ESR}_b\\ 
   \nonumber 
   SoC_{v,d,\rho}^{EV,RS}E^{C,EV}_b  - (1-X^{EV})~\overline{SoC}^{EV}_{v,d,\rho}~E^{C,EV}_b \nonumber \\\leq SoC^{C,RS,EV}_{v,d,\rho}\nonumber \\
   \leq  SoC_{v,d,\rho}^{EV,RS}~E^{C}_v  - (1-X^{EV})~\underline{SoC}^{EV}_{v,d,\rho}~E^{C,EV}_v
\end{gather}

\subsection{ESRs and EVs Market Constraints}
\label{constraint5}
This constraint of the ESR ensures that the ESR is only discharged within the minimum offered capacity.
\begin{gather}
    SoC_{b,j,t-1}^{ESR}-SoC^{ESR}_{b,j,t}  \leq \nonumber\\ \Gamma^{ESR,dch}_{b,j,t}~ \big (SoC^{ESR}_{b,j,t-1}-SoC^{ESR,RS}_{b,j,t-1}-\underline{SoC}^{ESR}_{b,j,t} \big ) \nonumber\\ 
    + (1-x^{ESR}_{b,j,t})E^{C,ESR}\\ 
        SoC_{v,j,t-1}^{EV}-SoC^{EV}_{v,j,t}  \leq \nonumber\\ \Gamma^{EV,dch}_{b,j,t}~ \big (SoC^{EV}_{v,j,t-1}-SoC^{EV,RS}_{v,j,t-1}-\underline{SoC}^{EV}_{v,j,t} \big ) \nonumber\\ 
    + (1-x^{EV}_{v,j,t} )E^{C,EV}
\end{gather}

\subsection{ESRs and EVs Operational Constraints}
\label{constraint6}
This constraint prohibits ESR/EV charging from the grid, maintains energy balance, and prevents discharge below the minimum state of charge (SoC) limit.

\begin{gather}
\small
    SoC_{b/v,j,t} = SoC_{b/v,t-1} - \frac{1}{\eta^{dch}} ~P_{b/v,t}
    \\ 
     SoC_{b/v} + SoC^{RS}_{b/v,t} \leq SoC_{b/v,t}\leq \overline{SoC}_{b/v,t}
     \\ 
     0 \leq P^{dch}_{b/v,t} \leq x \overline{P}^{dch}_{b/v,t} \\
     ~~~~\forall t \in \mathscr{T}, ~\forall b \in N_B, ~\forall v \in N_V, ~\forall x \in [1,0] \nonumber
\end{gather}








\subsubsection{Generator Voltage Constraints}


The generators' voltage magnitude limits are 
\begin{equation}
\small
V_{\text{min}} \leq V_i \leq V_{\text{max}}
\end{equation}
where $[V_{min,} V_{max}]=[0.95, 1.05]$. 
Additional constraints for generators, including line flows, voltage angle limits, and active and reactive power limits, are incorporated within the optimal power flow (OPF) algorithm.

The proposed event market model currently focuses on mobile generating resources for transmission network support during critical periods, excluding wind and solar sources. Future work will expand to detailed distribution level analysis and incorporate wind and solar plants into the modeling.

\begin{table}[]
\centering
\caption{Parameters of ESR and EVs.}
\label{table:parameters_ESR_EV}
\begin{tabular}{llll}
\hline
Parameters                 & Values & Parameters            & Values \\ \hline
SoC\_min                   & 0      & $\phi_{b}, \phi_{v}$  & -2.45  \\ \hline
SoC\_max                   & 1      & $\eta_{b}, \eta_{v}$  & 0.9    \\ \hline
$\alpha_{a}$, $\alpha_{v}$ & -36.23 & $\pi_x$ (\$/MWh/mile) & 3      \\ \hline
$\beta_{b}, \beta_{v} $    & 34.8   & $r$ (\%)              & 10     \\ \hline
$\gamma_{b}, \gamma_{v}$   & 2.77   &                       &        \\ \hline
\end{tabular}
\vspace{-18pt}
\end{table}

\vspace{-8pt}
\section{Numerical Simulation Results}
\subsection {Modified IEEE 30-bus Test System}
\label{caseStudy}
The resilience strategy for event markets proposed in this study is evaluated using the IEEE 30-bus system, with parameters sourced from MATPOWER~\cite{zimmerman2010matpower}. The test system, originally comprising 30 buses with six generator buses, is modified to include DGs, ESRs, and EVs for event scenarios. Specifically, DGs are installed at buses 4 and 9, ESRs at buses 6 and 21, and EVs at bus 8, with buses 4, 6, 8, 11, and 21 serving as generator buses. Generator costs are represented using piece-wise linear functions. Table \ref{table:parameters_ESR_EV} shows the resource parameters assumed for simulations. 

The study employs two scenarios to assess the proposed strategy, differentiated by event duration. In the first scenario, the event commences after the 5th hour and concludes with full system restoration by hour 12. The second scenario considers an event ending after the 9th hour. Both scenarios operate over a 12-hour optimization period with a 1-hour unit optimization time. This approach enables a comprehensive evaluation of the strategy's effectiveness in various event durations. The simulation conducts an OPF analysis of the network across three distinct phases to assess the system's behavior before, during, and after a high-impact event. \begin{inparaenum}
    \item \textbf{Phase 1 (Pre-Event Stage)}: This initial stage establishes the system's baseline state, providing crucial insights into its operational status prior to any event.
    \item \textbf{Phase 2 (Event Period)}: During this critical phase, the network undergoes stress by component outages, and OPF is used to identify resource deployment areas and the extent of these outages. This period allows operators to determine outage sizes and network areas needing reinforcement. From 1,000 initial outage scenarios, 10 are selected through k-means clustering. An event market is then activated based on the scenarios, leading to resource offer submissions. Post-market clearance, resources are allocated, and system's response is analyzed in detail.
    \item \textbf{Phase 3 (Post-Event Implementation)}: This final phase implements outcomes of the event market, integrating the allocated resource capacities into the network's operational framework.
\end{inparaenum}

\vspace{-10pt}
\subsection {Simulation Results}

\begin{figure}[!h]
\vspace{-18pt}
\centering
\subfloat[]{%
\centering
\includegraphics[clip,width=0.5\columnwidth]{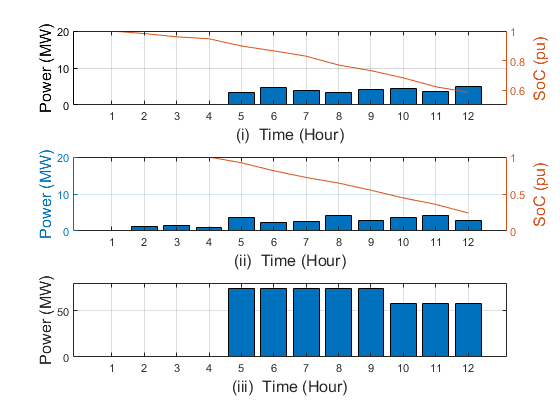}%
   \label{fig:power_case1_1}
}
\subfloat[]{%
\includegraphics[clip,width=0.5\columnwidth]{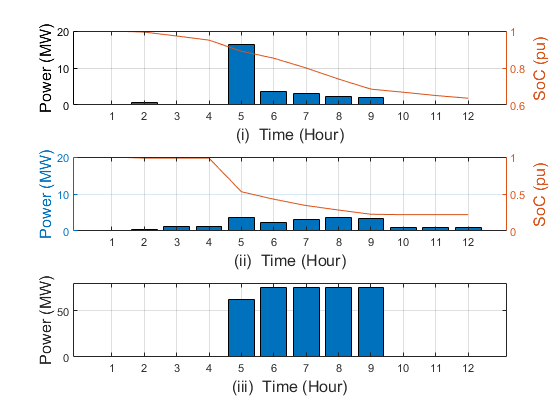}%
  \label{fig:power_case1_2}
}\\
\caption{ SoC \& power dispatched by each resource (a) Scenario 1 and (b) Scenario 2, (i) ESR, (ii) EV, and (iii) DGs.}
\label{fig:powerdisp30-118}
\vspace{-20pt}
\end{figure}


\begin{figure}[!h]
\centering
\subfloat[]{%
\centering
\includegraphics[clip,width=0.5\columnwidth]{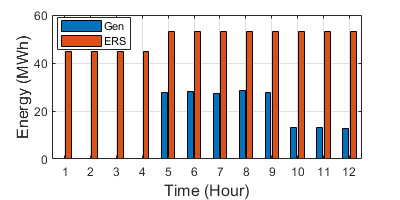}%
   \label{fig:reserve_case1_1}
}
\subfloat[]{%
\includegraphics[clip,width=0.5\columnwidth]{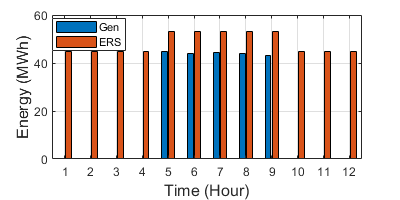}%
  \label{fig:reserve_case1_2}
}\\
\caption{ Total spinning reserve capacities by each resource (a) Scenario 1 and (b) Scenario 2.}
\label{fig:reserve30_118}
 \vspace{-12pt}
\end{figure}



\begin{figure}[!h]
\centering
\subfloat[]{%
\centering
  \includegraphics[width=4cm,height= 2.5cm]{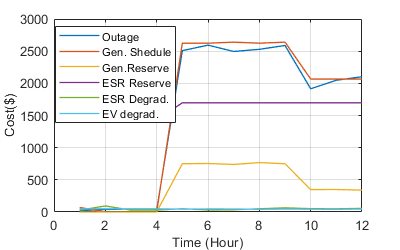}%
   \label{fig:cost1}
}
\subfloat[]{%
  \includegraphics[width=4cm,height= 2.5cm]{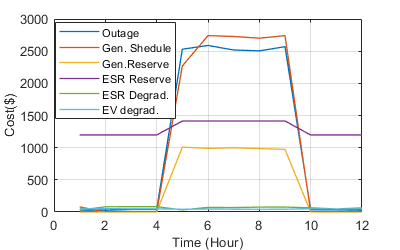}%
  \label{fig:cost2}
}\\

\caption{Total operation costs (a) Scenario 1  and (b) Scenario 2.}
\label{fig:totalcost}

 \vspace{-8pt}
\end{figure}

The outcomes of the event market are illustrated in Figs. \ref{fig:powerdisp30-118}--\ref{fig:totalcost}, covering Scenarios 1 and 2. Notably, the SoC for both ESRs and EVs showed a swift initial decline at the event's start, yet these resources remained active in grid operations post-event. The SoC levels suggest sufficient capacity availability for future system support if needed. If, for instance, the total SoC reaches zero before the event concludes, it indicates a shortfall in the system support from the event resources, impacting the network's resilience.

The total power discharged from (i) ESRs, (ii) EVs, and (iii) DGs during Scenarios 1 and 2 is detailed in Figs. \ref{fig:powerdisp30-118}(a) and (b), revealing a proportional response from each resource type relative to their capacities. It is observed that the output increases significantly from the given of the event. The dispatch is not entirely constant throughout the event, because other network resources (e.g., loads, other generators) also react to the new network conditions. Additionally, Fig.~\ref{fig:reserve30_118} illustrates the spinning reserve capacities, highlighting the ESRs' more active role compared to DGs in providing this reserve. Collectively, these results underscore a robust energy supply available to support the network during outages, with resources capable of contributing to the grid’s energy demands even after the event. This indicates not only the system's capacity to handle immediate challenges but also its resilience in maintaining reliable operations, thereby enhancing overall grid resilience.

Table \ref{table:market_quantities} offers a comprehensive breakdown of the market results, detailing various cost components for all resources in Scenarios 1 and 2. These include average values of ESR/EV or DG capacities, spinning reserve capacities cleared in the market, operational costs, degradation costs, spinning reserve costs, and transportation costs, along with the calculated social welfare. It is worth noting that the values obtained in Scenario 1 are higher than those in Scenario 2, except for travel distance costs. Consequently, more capacities are cleared for a longer event duration, leading to reduced social welfare. Additionally, Fig.~\ref{fig:totalcost} graphically presents how these costs fluctuate under different event conditions in both scenarios. While the degradation costs for ESRs/EVs remain constant throughout the operational period, the scheduling costs and reserve costs fluctuate based on the network condition and are proportional to the total capacities offered by each resource.

\begin{table}[!h]
\caption{Event market results for Scenarios 1 and 2.}
\label{table:market_quantities}
\begin{tabular}{ll|ll}
\hline
\multicolumn{1}{c}{\multirow{2}{*}{Quantity}}                             & \multicolumn{1}{c|}{Values} & \multicolumn{1}{c}{\multirow{2}{*}{Quantity}}                    & \multicolumn{1}{c}{Values} \\ \cline{2-2} \cline{4-4} 
\multicolumn{1}{c}{}                                                      & Scenario 1 (2)              & \multicolumn{1}{c}{}                                             & Scenario 1 (2)             \\ \hline
\begin{tabular}[c]{@{}l@{}}Avg. ESR \\ cleared (MWh)\end{tabular}         & 34.83 (21.43)               & \begin{tabular}[c]{@{}l@{}}ESR Reserve\\  cost (\$)\end{tabular}  & 19327 (15454)              \\ \hline
\begin{tabular}[c]{@{}l@{}}Avg. EV\\ cleared (MWh)\end{tabular}           & 39.51 (32.8)                & \begin{tabular}[c]{@{}l@{}}ESR deg.\\  cost (\$)\end{tabular}    & 555 (568)                  \\ \hline
\begin{tabular}[c]{@{}l@{}}Avg. DG\\  cleared (MW)\end{tabular}           & 45.39 (30.22)               & \begin{tabular}[c]{@{}l@{}}EV deg.\\  cost (\$)\end{tabular}     & 571 (779)                  \\ \hline
\begin{tabular}[c]{@{}l@{}}Avg. DG Reserve\\ cleared (MW)\end{tabular}    & 14.81 (15.29)               & \begin{tabular}[c]{@{}l@{}}DG transit\\  cost (\$)\end{tabular}  & 12960 (12960)              \\ \hline
\begin{tabular}[c]{@{}l@{}}Avg. ESR Reserve\\  cleared (MWh)\end{tabular} & 5.39 (4.31)                 & \begin{tabular}[c]{@{}l@{}}ESR transit\\  cost (\$)\end{tabular} & 28560 (28560)              \\ \hline
\begin{tabular}[c]{@{}l@{}}DG Sched.\\ cost (\$)\end{tabular}             & 19424 (13261)               & \begin{tabular}[c]{@{}l@{}}EV transit\\  cost (\$)\end{tabular}  & 278 (278)                  \\ \hline
\begin{tabular}[c]{@{}l@{}}DG Reserve\\  cost (\$)\end{tabular}            & 4800 (13261)                & Social Welfare                                                   & 65307 (61181)              \\ \hline
\end{tabular}
\end{table}
\vspace{-25pt}

\begin{figure}[!h]
\centering
\subfloat[]{%
\centering
  \includegraphics[width=0.43\linewidth]{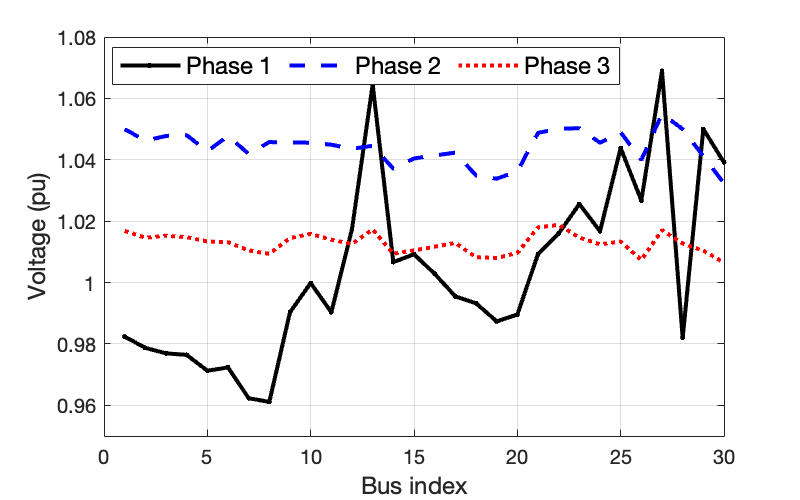}
 \label{fig:voltages30case1}
 }
 \subfloat[]{%
\centering
  \includegraphics[width=0.55\linewidth]{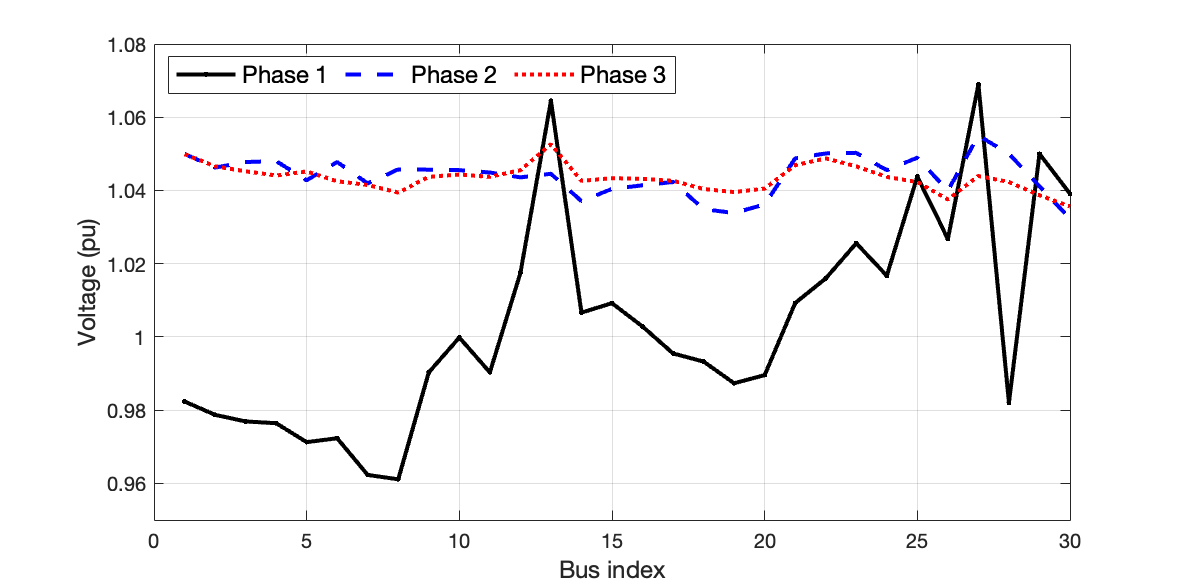}
 \label{fig:voltages30case2}
 }
 \caption{Voltage profile at the buses (a) Scenario 1   (b) Scenario 2.}
 \label{fig:voltages30bus}
\end{figure}

Fig.~\ref{fig:voltages30bus} displays the voltage profiles at various buses for Scenarios 1 and 2. In Phase 2 of both scenarios, the system experiences notable voltage fluctuations while determining outage size and resource allocation, pushing operational limits. However, in Phase 3, a stabilization in voltage levels is observed, aligning more closely with nominal values. Significantly, Scenario 1 exhibits a more distinct contrast in voltages between Phases 2 and 3, unlike Scenario 2, which shows less variation. This can be attributed to the difference in event duration and outage sizes between the scenarios, resulting in more pronounced operational changes. Despite these fluctuations, both scenarios successfully maintain voltages within the prescribed limits, underscoring the system's capability to adapt to varying event duration and stabilize post-event. This adaptability is key to ensuring continued grid resilience under varying stress conditions.

\section{Conclusion}
This paper presents an innovative event market model, specifically designed to deploy mobile generating systems efficiently in response to HILP events in the electric transmission network. Central to this approach is a GSM technique, crucial for forecasting the generation capacity needed near affected buses, demonstrated using the IEEE 30-bus system. Mobile generating resources, including ESRs and EVs, actively participate by offering specified quantities and prices. The event market's core function is to ascertain the precise supply capacities from each resource, vital for enhancing grid resilience. A thorough analysis of resource performance throughout operational phases has shown that this strategy offers a proactive, economically viable, and efficient means to mitigate the effects of such events on the grid. Key takeaways from this work include its emphasis on resilience, social welfare, and the importance of GSM in the event market.The strategy guarantees the provision of sufficient support to bolster the network and deliver benefits to all participants.

\bibliographystyle{IEEEtran}
\end{document}